\documentclass{article}
\usepackage{epsf,emulateapj}

\lefthead{MILLER, KNEZEK, \& BREGMAN}
\righthead{THE CLOSEST DAMPED LYMAN ALPHA SYSTEM}
\slugcomment{Accepted by Astrophysical Journal Letters}

\begin{document}

\def\SN{signal-to-noise ratio}
\def\NH{$N_{\rm H}$}
\def\mass{$M_{\odot}$}
\def\deg{^{\circ}}
\def\kps{km s$^{-1}$}
\def\etal{{\it et al.}}
\newcommand{\eex}[1]{\times 10^{#1}}
\newcommand{\ARAA}[2]{ARAA\rm, #1, #2}
\newcommand{\ApJ}[2]{ApJ\rm, #1, #2}
\newcommand{\ApJL}[2]{ApJ\rm, #1, L#2}
\newcommand{\ApJSS}[2]{ApJS\rm, #1, #2}
\newcommand{\AandA}[2]{A\&A\rm, #1, #2}
\newcommand{\AJ}[2]{AJ\rm, #1, #2}
\newcommand{\BAAS}[2]{BAAS\rm, #1, #2}
\newcommand{\ASP}[2]{ASP Conf. Ser.\rm, #1, #2}
\newcommand{\JCP}[2]{J. Comp. Phys.\rm, #1, #2}
\newcommand{\MNRAS}[2]{MNRAS\rm, #1, #2}
\newcommand{\N}[2]{Nature\rm, #1, #2}
\newcommand{\PASJ}[2]{PASJ\rm, #1, #2}
\newcommand{\PASP}[2]{PASP\rm, #1, #2}
\newcommand{\RPP}[2]{Rep. Prog. Phys.\rm, #1, #2}

\title{The Closest Damped Lyman Alpha System}
\author{Eric D.\ Miller}
\affil{Dept.\ of Astronomy, University of Michigan,
Ann Arbor, MI 48109-1090 
emiller@astro.lsa.umich.edu}
\author{Patricia M.\ Knezek}
\affil{Dept. of Physics \& Astronomy, The Johns Hopkins University,
Baltimore, MD 21218-2695 
pmk@pha.jhu.edu}
\author{Joel N.\ Bregman}
\affil{Dept.\ of Astronomy, University of Michigan,
Ann Arbor, MI 48109-1090 
jbregman@umich.edu}

\begin{abstract}

A difficulty of studying damped Ly$\alpha$ systems is that they are
distant, so one knows little about the interstellar medium of the galaxy.
Here we report upon a damped Ly$\alpha$ system in the nearby galaxy NGC
4203, which is so close $(v_{helio}=1117$ \kps) and bright ($B_{o}=11.62$)
that its \ion{H}{1} disk has been mapped.  The absorption lines are
detected against Ton 1480, which lies only 1.9$^{\prime}$\ (12 $h_{50}$
kpc) from the center of NGC 4203.  Observations were obtained with the
Faint Object Spectrograph on HST (G270H grating) over the 2222--3277 \AA\
region with 200 \kps\ resolution.  Low ionization lines of Fe, Mn, and Mg
were detected, leading to metallicities of $-2.29$, $< -0.68$, and $>
-2.4$, which are typical of other damped Ly$\alpha$ systems, but well below
the stellar metallicity of this type of galaxy.  Most notably, the velocity
of the lines is 1160 $\pm$ 10 \kps, which is identical to the \ion{H}{1}
rotational velocity of 1170 \kps\ at that location in NGC 4203, supporting
the view that these absorption line systems can be associated with the
rotating disks of galaxies.  In addition, the line widths of the Mg lines
give an upper limit to the velocity dispersion of 167 \kps\, to the 99\%
confidence level.

\end{abstract}


\section{Introduction}

Damped Ly$\alpha$ systems (hereafter, DLA) are the primary tracers of
metallicity over cosmological distances ({\it e.g.}, Lu \etal\ 1996) and
provide a method of identifying distant galaxies ({\it e.g.}, Giavalisco,
Steidel, and Szalay 1994). It is likely that these DLA are associated with
galaxies, but whether the lines arise in a rotating disk or in the halo
remains a matter of controversy (Charlton and Churchill 1996; Prochaska and
Wolfe 1997; Lu, Sargent, and Barlow 1997). One reason for this uncertainty
is that we possess little information on the cold gas content of these
galaxies. \ion{H}{1} velocity maps of the absorbing hosts would be very
valuable in addressing this issue, but even the two previously known
nearest DLAs, at $z=0.101$ and $z=0.167$ (Petitjean \etal\ 1996; Lanzetta
\etal\ 1997), are too distant for such observations to be feasible.

Another issue is the meaning of the measured abundances, which nominally
indicate that the metallicities of DLA are typically 10--300 times below
the solar value over redshifts of $z =$ 0--3.  (For higher redshifts, the
metallicity appears to be even lower; Lu, Sargent, and Barlow 1998.)
However, low apparent abundances are also derived for sight lines out of
the Milky Way, but in this case we know this is due to depletion onto
grains ({\it e.g.}, Cardelli, Sembach, and Savage 1995; Savage and Sembach
1996). Unfortunately, depletion onto grains can raise a substantial
uncertainty in the metallicity deduced from DLAs.

Some insight into these issues is offered by our observations of a DLA
caused by a galaxy so nearby that some of the properties of its \ion{H}{1}
disk are already known. Here we present ultraviolet resonance line
observations that led to the discovery of a DLA system toward the quasar
Ton 1480 ($z = 0.614$; $V = 16.5$), which lies behind the nearby galaxy NGC
4203 at a projected separation of only 1.9$^{\prime}$\ (12 $h_{50}$\ kpc)
from the center. NGC 4203 is a nearby $(v_{helio}=1117$\ \kps), bright
$(B_{o}=11.62)$ early-type galaxy that is not in a cluster, has normal
colors, with $D_{25}=3.6^{\prime}$, and a one-dimensional velocity
dispersion of 175 \kps. The background quasar, Ton 1480, was discovered by
us as part of a program whereby we use background X-ray sources to identify
quasars near targets of interest, such as galaxies and galaxy clusters
(Knezek and Bregman 1998). This galaxy has been observed in many wavebands,
including an \ion{H}{1} map (van Driel \etal\ 1988) with both column
density and velocity information for the neutral gas.  In the following, we
discuss the absorption line observations, the column densities for Fe, Mg,
and Mn, the kinematics of the absorbing gas, and how this relates to the
known kinematics of the \ion{H}{1} gas.

\section{Data}

The data were taken in January 1997 with the Faint Object Spectrograph
aboard HST.  The 1640 second exposure was taken through the
0.86$^{\prime\prime}$\ aperture using the G270H grating and Red Digicon
detector.  In this configuration, the expected FOS instrumental line width
(FWHM) is 1.97 \AA, providing a spectral resolution of 211 \kps\ at the
\ion{Mg}{2} $\lambda$2800 doublet.  The wavelength coverage ranges from
2222 to 3277 \AA,  with a S/N ratio of about 20 throughout the spectrum.

Visual inspection of the spectrum reveals several strong absorption
features (see Figure \ref{fig1}).  The \ion{Mg}{2} $\lambda$2800 doublet
from NGC 4203 can be seen shifted 10.8 \AA\ (1160 \kps) from its Galactic
counterpart.  The \ion{Fe}{2} $\lambda$2600 line also can be seen shifted
10.1 \AA\ (1165 \kps) from the Galactic line.  In general, lines with
$W_{\lambda} > 0.5$ \AA\ are easily visible to the eye.  Other such lines
include \ion{Fe}{2} $\lambda$2374, $\lambda$2382, as well as their
corresponding Galactic lines.

To detect fainter absorption features, we utilized the line-searching
software developed for the HST absorption line key project (Schneider
\etal\ 1993).  This algorithm performs both Gaussian and PSF fitting to
absorption features across the spectrum, finding line centers and
equivalent widths with errors.  The detection threshold was set to
3-$\sigma$, which corresponds to a minimum equivalent width of 0.15 \AA\
across most of the spectrum.  The fitting procedure detected 20 lines (see
Table \ref{tab:data}); of these, 7 are Galactic in origin, 6 are from the
DLA, 3 are blends of Milky Way and DLA lines, and 4 are unidentified.  For
a spectrum with 2064 resolution elements, we would expect roughly 1
spurious detection above the 3-$\sigma$\ limit.  Thus a few of the unknown
lines, which are all near the 3-$\sigma$\ limit, could possibly be real
lines; however, they are inconsistent with any known atomic resonance
lines, either Galactic or at the velocity of NGC 4203.  It remains to be
seen if they are real absorption lines from intervening material between
the quasar and NGC 4203.

We identified weak lines in \ion{Fe}{2}, \ion{Mg}{1}, and \ion{Mn}{2} at
the redshift of NGC 4203, as well as the easily visible lines mentioned
above.  All identified DLA lines are listed in Table \ref{tab:data} which
gives the vacuum rest wavelength, line center, and equivalent width with
1-$\sigma$\ errors.

Analysis of the Galactic features sheds some light on the quality of the
HST/FOS wavelength calibration.  At the high Galactic latitude of Ton 1480
($l=173\deg$, $b=80.1\deg$), the Galactic ISM should be nearly at rest
relative to the heliocentric velocity.  Thus any Galactic line should be
barely shifted from its rest wavelength in the direction of Ton 1480.

To determine the average expected shift in the Milky Way lines, we
calculated a weighted mean line-of-sight velocity for the \ion{H}{1} disk,

\[ <v_{\rm LOS}> = \frac 
{{\displaystyle \int\!\!\int} v_{\rm LOS}(R,z) n(R,z) dR dz} 
{{\displaystyle \int\!\!\int} n(R,z) dR dz} , \]

\noindent integrating out to $z=500$ pc.  Using the estimate of the
\ion{H}{1} density $z$-dependence from Dickey and Lockman (1990) and the
standard equation for differential Galactic rotation, $v_{\rm
LOS}=\Theta_{\odot} [ R_{\circ}/R -1 ] \sin{l} \cos{b}$, we find $<v_{\rm
LOS}>=-4$\ \kps, only 2\% of our spectral resolution.

From the FOS calibrated (vacuum) wavelengths, the average shift of the
clean Galactic features is $-1.23 \pm 0.2$ \AA.  This corresponds to a
shift of $-130 \pm 20$\ \kps\ at 2800 \AA, much larger than the expected
shift found above.  Thus the FOS wavelength calibration is off by about 1.2
\AA\ to the blue.  This is within the pointing error for a FOS Binary
Target Acquisition; in addition, it has been found recently that the
zero-point of the FOS wavelength scale varied considerably over the
lifetime of the instrument (Keyes 1998).  Since we did not obtain a lamp
spectrum at the time of the observation but instead relied on the default
calibration used in the HST pipeline, our wavelength calibration could
indeed be off by this amount.

\centerline{\null}
\vskip2.65truein
\includegraphics{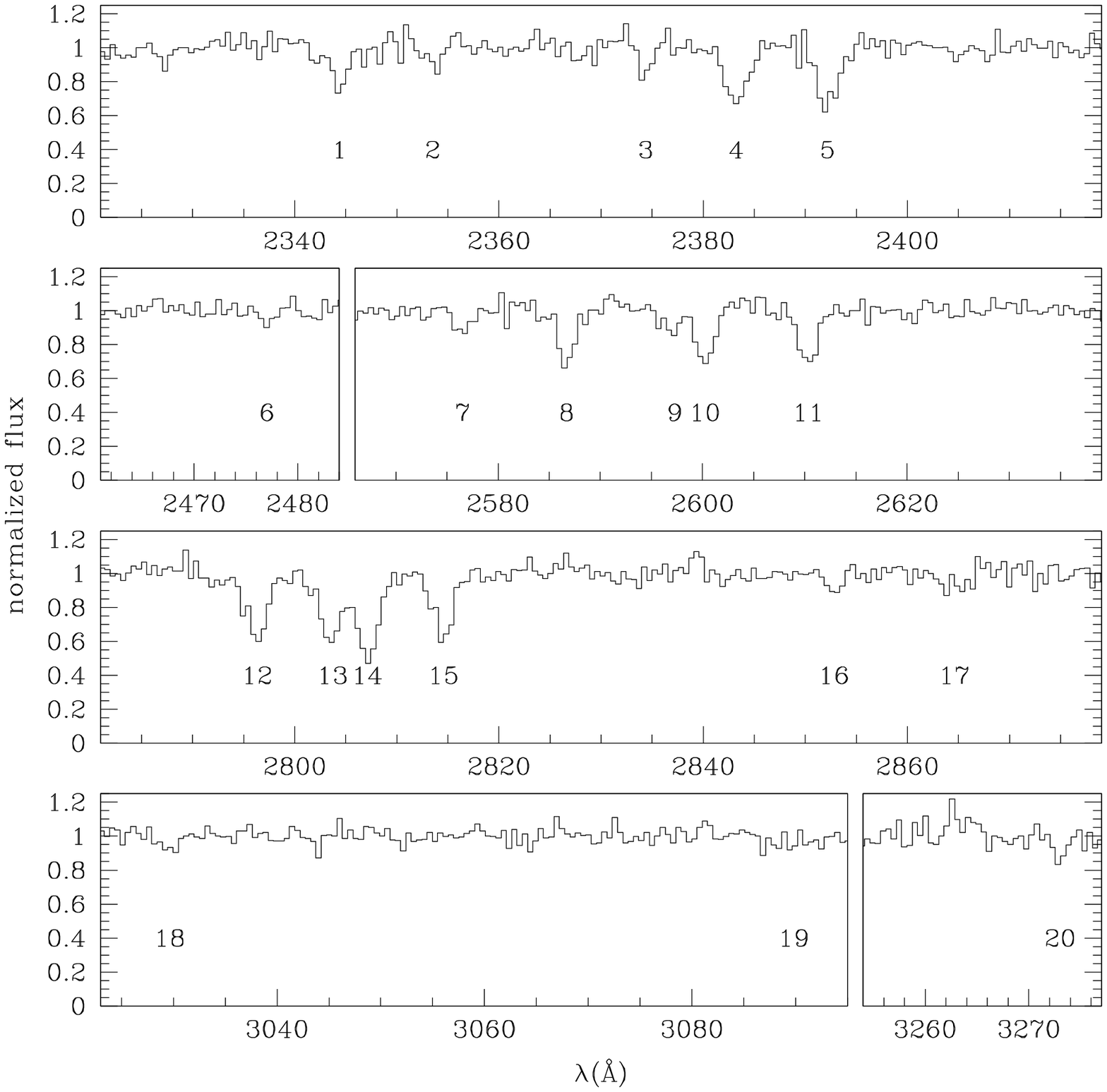}
\figcaption{
Selected regions of the normalized spectrum of Ton 1480.  The
labels correspond to ID entries of absorption lines in Table \ref{tab:data}.
\label{fig1}}

\vskip0.1truein

Consequently, we used the Galactic lines (plus the 4 \kps\ offset) as the
fiducial zero velocity. The velocity shifts in all DLA lines from the
Galactic zero-point are given in Table \ref{tab:data}.  (For display
purposes, all wavelength measurements have been shifted 1.2 \AA\ to the
red.)  The average line shift for the unblended lines is $1163 \pm 10$\
\kps\ from the Galactic reference lines, implying a heliocentric velocity
of $v_{\rm helio} = 1159 \pm 10$\ \kps.

As mentioned earlier, several absorption lines are blended with Milky Way
features.  The \ion{Fe}{2} $\lambda$2374, \ion{Fe}{2} $\lambda$2587 and
\ion{Mn}{2} $\lambda$2577 lines are all blended with Galactic \ion{Fe}{2}
lines.  This \ion{Mn}{2} line is the only line of this element detected, so
only an upper limit can be obtained for its abundance.  For Fe, three lines
(\ion{Fe}{2} $\lambda$2344, \ion{Fe}{2} $\lambda$2383, and \ion{Fe}{2}
$\lambda$2600) were cleanly detected.  Additionally, non-detections of two
\ion{Fe}{2} lines and three \ion{Fe}{1} lines provide upper limits for
their equivalent widths.  Finally, we have detected three Mg lines in two
stages of ionization, the \ion{Mg}{2} $\lambda$2800 doublet and \ion{Mg}{1}
$\lambda$2853.  The doublet is clean and pronounced while the \ion{Mg}{1}
line is weak and appears broadened.  It also has an anomalous velocity,
shifted 75 \kps\ from the mean determined above.  It is possible this line
is blended, although there are no strong Galactic features in this region
of the spectrum.  Thus its nature remains questionable, and we treat it
only as an upper limit.

\begin{table*}[t]
\caption[NGC 4203 Absorption Line Data]{}
\label{tab:data}
\begin{center}
\begin{tabular}{lcclcrr}
\multicolumn{7}{c}{\sc NGC 4203 absorption line data} \cr
\tableline \tableline
Species & ID\tablenotemark{a} & $\lambda$(\AA) & $\lambda _{obs}$(\AA) &
	$v_r$(km/s) & $W_{\lambda}$(\AA) & $N$(cm$^{-2}$) \\
\tableline
\ion{Fe}{1}  & --- & 2484.02 & 2494.5 & ---
     & $\leq$\ 0.091 & $\leq 3.0 \eex{12}$ \nl 
\ion{Fe}{1}  & --- & 2523.61 & 2533.3 & ---
     & $\leq$\ 0.082 & $\leq 5.2 \eex{12}$ \nl 
\ion{Fe}{1}  & --- & 2719.83 & 2730.4 & ---
     & $\leq$\ 0.104 & $\leq 1.3 \eex{13}$ \nl \tableline
\ion{Fe}{2} & 2 & 2344.21 & 2353.5$\pm$0.4 & 1168.9 
     & 0.280$\pm$0.064 & 5.3$\pm 1.2 \eex{13}$ \nl 
\ion{Fe}{2} & 5 & 2382.77 & 2392.2$\pm$0.1 & 1124.8\tablenotemark{b}
     & 0.804$\pm$0.049 & 5.3$\pm 0.3 \eex{13}$ \nl 
\ion{Fe}{2} & 11 & 2600.17 & 2610.3$\pm$0.1 & 1164.5 
     & 0.683$\pm$0.041 & 5.1$\pm 0.3 \eex{13}$ \nl 
\ion{Fe}{2} & --- & 2249.88 & 2259.0 & ---
     & $\leq$\ 0.129 & $\leq 1.1 \eex{15}$ \nl 
\ion{Fe}{2} & --- & 2260.78 & 2270.7 & ---
     & $\leq$\ 0.127 & $\leq 7.6 \eex{14}$ \nl 
\ion{Fe}{2} & 4 & 2374.46 & 2383.2$\pm$0.1 & 1117.4\tablenotemark{c}
     & $\leq$\ 0.773\tablenotemark{c} & $\leq 5.5 \eex{14}$ \nl 
\ion{Fe}{2} & 9 & 2586.65 & 2597.3$\pm$0.2 & 1225.1\tablenotemark{b,c}
     & $\leq$\ 0.281\tablenotemark{c} & $\leq 7.4 \eex{13}$ \nl \tableline
\ion{Mg}{1}  & 17 & 2852.96 & 2864.6$\pm$0.3 & 1238.9 
     & 0.203$\pm$0.052 & $\leq 1.5 \eex{12}$\tablenotemark{d} \nl \tableline
\ion{Mg}{2} & 14 & 2796.35 & 2807.1$\pm$0.1 & 1156.8 
     & 1.210$\pm$0.039 & $\geq 2.9 \eex{13}$ \nl 
\ion{Mg}{2} & 15 & 2803.53 & 2814.5$\pm$0.1 & 1162.4 
     & 0.857$\pm$0.042 & $\geq 4.0 \eex{13}$ \nl \tableline
\ion{Mn}{2} & 8 & 2576.88 & 2586.7$\pm$0.1 & 1180.8\tablenotemark{c}
     & $\leq$\ 0.649\tablenotemark{c} & $\leq 3.1 \eex{13}$ \nl 
\ion{Mn}{2} & --- & 2594.50 & 2604.6 & ---
     & $\leq$\ 0.097 & $\leq 6.0 \eex{12}$ \nl 
\ion{Mn}{2} & --- & 2606.46 & 2616.7 & ---
     & $\leq$\ 0.098 & $\leq 8.5 \eex{12}$ \nl \tableline
Gal \ion{Fe}{2} & 1 & 2344.21 & 2344.4$\pm$0.2 & ---
     & 0.493$\pm$0.056 & ---  \nl
Gal \ion{Fe}{2} & 3 & 2374.46 & 2374.4$\pm$0.3 & ---
     & 0.258$\pm$0.059 & ---  \nl
Gal \ion{Mn}{2} & 7 & 2576.88 & 2576.6$\pm$0.2 & ---
     & 0.270$\pm$0.044 & ---  \nl
Gal \ion{Fe}{2} & 10 & 2600.17 & 2600.2$\pm$0.1 & ---
     & 0.675$\pm$0.041 & ---  \nl
Gal \ion{Mg}{2} & 12 & 2796.35 & 2796.3$\pm$0.1 & ---
     & 1.000$\pm$0.063 & ---  \nl
Gal \ion{Mg}{2} & 13 & 2803.53 & 2803.7$\pm$0.1 & ---
     & 1.060$\pm$0.072 & ---  \nl
Gal \ion{Mg}{1} & 16 & 2852.96 & 2852.8$\pm$0.3 & ---
     & 0.224$\pm$0.051 & ---  \nl
unknown & 6 & --- & 2477.0$\pm$0.4 & ---
     & 0.140$\pm$0.045 & ---  \nl
unknown & 18 & --- & 3029.6$\pm$0.4 & ---
     & 0.179$\pm$0.050 & ---  \nl
unknown & 19 & --- & 3089.8$\pm$0.4 & ---
     & 0.287$\pm$0.095 & ---  \nl
unknown & 20 & --- & 3273.0$\pm$0.4 & ---
     & 0.250$\pm$0.084 & ---  \nl
\tableline
\tablenotetext{a}{Absence of a line ID indicates a non-detection.}
\tablenotetext{b}{The Galactic counterparts of these lines are blended, so
it is difficult to determine an accurate velocity.}
\tablenotetext{c}{These DLA lines are blended, leading to inaccurate
velocities and upper limits on $W_{\lambda}$\ and $N$.}
\tablenotetext{d}{The anomalous velocity of the \ion{Mg}{1} line makes its
nature questionable, thus we include it as an upper limit.}
\end{tabular}
\end{center}
\end{table*}

\section{Interpretation \& Conclusions}

Column densities can be determined for weak (optically thin) lines by using
the relation ({\it e.g.}, Spitzer 1978):

\[ N = 1.13 \eex{20} W_{\lambda} ({\rm \AA}) / 
   f \lambda^{2} ({\rm \AA})~~{\rm cm}^{-2}\]

\noindent
The unblended \ion{Fe}{2} lines are all thin lines, since the value of
$W_{\lambda} / \lambda^{2}$ scales linearly with the oscillator strength,
$f$ (taken from Morton 1991).  The \ion{Mg}{2} $\lambda$2800 doublet lines
appear to be on the Doppler portion of the curve of growth, since the line
strength does not scale as the oscillator strength.  Since we have two
lines, we can simply take the line ratio, and we find that \(
W_{\lambda2803} / W_{\lambda2796} = 0.71 \pm 0.05. \) In the thin-line
limit, we would expect the line ratio to be \( W_{\lambda2803} /
W_{\lambda2796} = (f \lambda^2)_{\lambda2803} / (f \lambda^2)_{\lambda2796}
= 0.50, \) so these lines are indeed optically thick, and we can only
obtain a lower limit to the \ion{Mg}{2} column density.  The \ion{Mg}{1}
and \ion{Mn}{2} lines are probably thin, but it is problematic to determine
column densities without more lines from these species.

Column densities and 1-$\sigma$\ errors for each line are given in Table
\ref{tab:data}.  For lines which were not detected, a 3-$\sigma$\ upper
limit to the equivalent width can be determined from consideration of the
detection threshold and noise.  Assuming the lines are weak, these values
represent an upper limit to the column density.

To obtain the abundance of an element relative to solar, an accurate
measurement of the \ion{H}{1} column is needed.  For this system, there are
several possible methods.  Van Driel \etal\ (1988) used 21-cm observations,
finding an \ion{H}{1} column of about $1 \eex{20}$\ cm$^{-2}$ towards Ton
1480.  A more accurate estimate comes from X-ray absorption observations.
The X-ray absorption column toward Ton 1480 was determined for the
\textit{ROSAT PSPC} data, which yielded a value of 3.38$ \eex{20}$\
cm$^{-2}$ (Bregman, Hogg, and Roberts 1995), while the Galactic \ion{H}{1}
column in this direction is only 1.14$ \eex{20}$\ cm$^{-2}$ (Hartmann and
Burton 1997).  Since the value of the Galactic \ion{H}{1} column is the
same as the Galactic X-ray absorption column (Arabadjis and Bregman 1998,
and references therein), we attribute the excess absorption column to that
of the disk of NGC 4203, which leads to a value of \NH $=2.2\eex{20}$\
cm$^{-2}$.  To further constrain this value, Ly$\alpha$ observations are
planned for HST in the near future.

To obtain the total elemental abundance, we need to know the column density
in each stage of ionization.  We are severely limited since we have
detected only one ionization stage for two of the three elements.  Thus we
need to model the ionization structure to correct for this.  The ionization
model in this region of NGC 4203 was produced using CLOUDY (Ferland et al.\
1998).  Two different ionizing spectra (Ikeuchi and Ostriker 1986; Haardt
and Madau 1996) were tested, along with the two values for the \ion{H}{1}
column \NH\ given above.  For most of our lines, there is little difference
in the ionization among the four models.  \ion{Fe}{1} and \ion{Fe}{2}
account for $\sim$ 5\% and $\sim$ 90\% of the Fe in each model,
respectively.  The fractional ionization of Mg is \ion{Mg}{1} $\sim$
10--20\%\ and \ion{Mg}{2} $\sim$ 70--75\%.  It should be noted that
adjusting \NH\ by a factor of 2 makes less of a difference than changing
the spectrum of the ionizing radiation.  This is especially true for
\ion{Fe}{1}, where the difference between the two model spectra is a factor
of 2 in the fractional ionization.  

We used the spectrum from Figure 5a of Haardt and Madau (1996), normalized
to $J_{\nu}$(912 \AA) $= 2\eex{-23}$\ ergs cm$^{-2}$ s$^{-1}$ Hz$^{-1}$
sr$^{-1}$ (Dove and Shull 1994), and the X-ray column of \NH $=2.2\eex{20}$\
cm$^{-2}$. The model gives the following ionization fractions: \ion{Fe}{1}
= 0.038, \ion{Fe}{2} = 0.942, \ion{Mg}{1} = 0.117, \ion{Mg}{2} = 0.764, and
\ion{Mn}{2} = 0.635.  The relative column densities for \ion{Fe}{1} and
\ion{Fe}{2} are consistent with this model.  However, according to our data
$N({\rm Mg}^0)/N({\rm Mg}^+) = 0.04$, whereas our model predicts it to be
0.15.  It appears the ionization in this region is higher than our model
predicts; this is especially likely given the anomalous velocity of the
\ion{Mg}{1} line, which could indicate that the \ion{Mg}{1} and \ion{Mg}{2}
absorbers are not co-spatial.  

Earlier results indicate that ionization correction factors for \ion{Mg}{2}
and \ion{Fe}{2} may be insignificant for this observed \ion{H}{1} column
density (Viegas 1995; Prochaska and Wolfe 1996).  We find our data to be
consistent with these results, and so we will not apply any ionization
correction to these species for the following analysis.

The column densities listed in Table \ref{tab:data} give the following
results for the heavy element abundances, relative to the solar values of
$log({\rm Fe}) = 7.67$, $log({\rm Mn}) = 5.39$, $log({\rm Mg}) = 7.58$, and
$log({\rm H}) = 12.00$ (Anders and Grevesse 1989):

\noindent {[}Fe/H] $= -2.29 \pm 0.10$  \\
{[}Mn/H] $< -0.68 $  \\
{[}Mg/H] $> -2.4 $  

These values are significantly below the expected stellar metallicity,
although it is difficult to separate true underabundances from depletion
onto grains since each of these elements have high condensation
temperatures.

A comparison between the known velocity of the \ion{H}{1} disk with the
absorption line velocities can help us understand the nature of the
absorption site.  Van Driel's (1988) velocity maps give an \ion{H}{1}
velocity of 1170 \kps\ along this line of sight in NGC 4203.  From the
centers of the unblended \ion{Fe}{2} and \ion{Mg}{2} absorption lines, the
average heliocentric velocity is found to be 1160 $\pm$ 10 \kps.  This is
indistinguishable from the \ion{H}{1} value, clearly demonstrating that the
absorption is due to a galactic disk of gas.

Limits can also be placed on the internal motions of the absorbing gas.
The widths of the strongest \ion{Fe}{2} lines average to $260 \pm 27$ \kps\
FWHM, with an instrumental line width of 241 \kps, which sets an upper
limit to the velocity dispersion of 224 \kps\ at the 99\% confidence level.
For \ion{Mg}{2}, the constraint is stricter, with an average width of $248
\pm 12$ \kps\ FWHM and an instrumental profile of 210 \kps\ yielding a
maximum velocity dispersion of 167 \kps\ at 99\% confidence.  

In the future, we hope to use this unique system to pursue additional
issues of DLAs.  A planned HST observation with STIS should determine
whether high ionization state gas is present, permit column densities to be
measured for other species, and provide a third measure of the \ion{H}{1}
column.  Higher spectral resolution studies will be able to examine the
velocity structure of the absorbing material.  Finally, the intrinsic
metallicity of the system can be studied through optical observations of
\ion{H}{2} regions and by X-ray studies.

We would like to thank J. Bergeron, J. Charlton, D. Schneider, C. Cowley,
E. Schulman, and the staff at STScI for their advice and assistance.
Financial support for this work has been provided by NASA through NAG5-3247
and STSCI GO-06547.


\end{document}